# The He$_2$ - OCS complex: comparison between theory and experiment

J. Norooz Oliaee,[1] N. Moazzen-Ahmadi,[1] A.R.W. McKellar,[2] Xiao-Gang Wang,[3] and Tucker Carrington, Jr.[3]

[1] *Department of Physics and Astronomy, University of Calgary, 2500 University Drive North West, Calgary, Alberta T2N 1N4, Canada.*

[2] *National Research Council of Canada, Ottawa, Ontario K1A 0R6, Canada.*

[3] *Chemistry Department, Queen's University, Kingston, Ontario K7L 3N6, Canada.*

**Abstract**

OCS is an ideal probe for quantum solvation effects in cold helium clusters. He$_2$-OCS is the "second step" in going from a single OCS molecule to a large doped superfluid helium cluster. Here assignment of the spectrum of He$_2$-OCS is significantly extended to higher values of $J$, $K$, and $v_t$ (the low frequency torsional vibration). The observation of a new infrared band, OCS $v_1 + v_3$, taken together with the known $v_1$ band, enables assignments to be verified by comparing ground state combination differences. Relatively straightforward scaling of previously calculated theoretical energy levels gives a remarkably good fit to experiment.

Suggested running head: He$_2$-OCS

Address for correspondence: Prof. N. Moazzen-Ahmadi,
    Department of Physics and Astronomy,
    University of Calgary,
    2500 University Drive North West,
    Calgary, Alberta T2N 1N4
    Canada.

**1. Introduction**

The OCS molecule has proven to be a valuable probe of "quantum solvation" and microscopic superfluid effects in helium clusters and droplets [1 - 15], and therefore the He – OCS interaction potential is of some interest [16 - 23]. The first step in this solvation is represented by the binary van der Waals complex He-OCS, whose spectrum has been studied in detail in the microwave [16, 24] and infrared [25 - 28] regions.* The second step in solvation is then the He$_2$-OCS complex, which is the subject of the present paper.

Limited experimental results on He$_2$-OCS from microwave [7,29] and infrared [6,27] spectra have been published previously. As well, detailed theoretical studies were published in 2010 by Wang and Carrington [30] and by Li et al. [31]. He-OCS has a T-shaped structure, with the He atom lying to the "side" of the OCS molecule. The center of mass separation is about 3.83 Å, and the effective angle between the O-C-S molecular axis and the line connecting the centers of mass is about 66°. When a second helium is added to make He$_2$-OCS, it occupies an analogous position, but displaced azimuthally by roughly 4 Å from the original helium atom, giving a dihedral structure. However, there is only a low energy barrier for the He atoms to tunnel to the "other side" of the OCS.

The situation is similar to the previously-studied cases of He$_2$-Cl [32,33], He$_2$-CO [34], He$_2$-CO$_2$ [35], and He$_2$-N$_2$O [36]. In all these cases, there are very low energy vibrational states which can be characterized by a torsional quantum number, $v_t$. In the present case of He$_2$-OCS, the first and second excited torsional vibrations $v_t = 1$ and 2 have calculated energies [30] only 0.43 and 1.58 cm$^{-1}$ above the ground state, $v_t = 0$. The calculated $v_t = 1$ and 2 energies are quite similar for $He_2$-$CO_2$, 0.50 and 1.66 cm$^{-1}$, and for $He_2$-$N_2O$, 0.60 and 1.87 cm$^{-1}$ [35,36]. The $v_t$ states have different symmetry properties, and it turns out that only even values of the rotational quantum number $K_a$ are possible for even values of $v_t$, and only odd $K_a$ for odd $v_t$, due to the effects of nuclear spin symmetry and the zero spin of $^4$He.

---

* Note there is a typo for one parameter in Table I of Ref. [27]: the value of $H_K'$ for He-$^{16}$O$^{12}$C$^{32}$S should be ten times smaller than listed.

In previous work on $He_2$-OCS, it was relatively easy to assign the transitions involving $(v_t, K_a) = (0, 0)$ levels. Some $(v_t, K_a) = (1, 1)$ and $(0, 2)$ transitions were also assigned [30], but these were more problematical because the differences between calculated and observed line positions were sometimes similar in magnitude to the differences among various alternate assignments. This would not be such a problem if there was a simple empirical Hamiltonian to fit the observed $He_2$-OCS spectrum without too many free parameters. But it turns out that different $v_t$ states have quite different rotational parameters, and that in any case a normal asymmetric rotor Hamiltonian does not work particularly well for the theoretical levels [30], as shown below.

Previous infrared study of He-OCS was limited to the region of the strong OCS $v_1$ fundamental band ($\approx$2060 cm$^{-1}$), but we recently extended it to the much weaker OCS $v_1 + v_3$ combination band ($\approx$2918 cm$^{-1}$) [28]. The same $v_1 + v_3$ spectrum also contains $He_2$-OCS transitions, which now provide the opportunity to use ground state combination differences to confirm (or otherwise) the $K_a = 1$ and 2 transitions mentioned above, since the $v_1$ and $v_1 + v_3$ bands share a common lower state. Thus in the present paper, we report an extension of the assignments in the experimental spectra of $He_2$-OCS. Highlights include extension of the existing $(v_t, K_a) = (0, 0)$, $(0, 2)$, and $(1, 1)$ assignments to higher $J$-values, and completely new assignments of $(v_t, K_a) = (0, 2 \leftrightarrow 0)$, $(1, 3)$ and $(2, 0)$ transitions. In addition to the normal isotopologue, $He_2$-$^{16}O^{12}C^{32}S$, we also include results for $He_2$-$^{16}O^{13}C^{32}S$ and $He_2$-$^{16}O^{12}C^{34}S$. The many new assignments enable construction of partial tables of $He_2$-OCS rotational energy levels which are "experimental", that is, not reliant on theory or on a particular energy formula or Hamiltonian.

## 2. Theoretical rotational levels

Details of the calculations, which utilized the *ab initio* He-OCS potential surface of Paesani and Whaley [21], were given previously [30]. The resulting energy levels for the ground vibrational state of $He_2$-OCS are given in Table A-1 (Supplementary Data) for the $v_t = 0$, 1, and 2 torsional states with $J$ values of 0 to 5. Even though we know that odd $K_a$ values are forbidden for even $v_t$ values (and even $K_a$ for odd $v_t$), they can still be calculated and are given here. The theoretical paper on $He_2$-OCS by Li et al. [31] focused on the $v_t = 0$, $K_a = 0$ levels for numerous isotopologues. Their results are in good general agreement with those of [30], but differ in detail because a different He-OCS potential [37] was used.

Table 1. Asymmetric rotor fits to $He_2$ – OCS theoretical [30] energy levels (in cm$^{-1}$).

|  | $v_t = 0$ | $v_t = 1$ | $v_t = 2$ |
| --- | --- | --- | --- |
| $T_0$ | 0.0000 | 0.4296 | 1.5845 |
| $A$ | 0.1891(4) | 0.2012(4) | 0.1925(4) |
| $B$ | 0.1489(4) | 0.1416(14) | 0.1423(5) |
| $C$ | 0.1021(4) | 0.1164(14) | 0.1078(4) |
| rms error | 0.0125 | 0.0150 | 0.0091 |

It is interesting to see how well the theoretical levels can be fitted with a conventional asymmetric rotor Hamiltonian. Results are shown in Table 1 for such fits, which were made using three rotational and five quadratic centrifugal distortion parameters to fit the theoretical levels for $J = 1$ to 4, including forbidden ones. (See Supplementary Data for the distortion parameters.) As a test, we could then see how well the $J = 5$ levels were predicted. The overall root mean square errors for $J = 1$ to 5 are given at the bottom of Table 1, and their rather large values show that the asymmetric rotor fits were not particularly successful. Of course, the fits could be improved by including more distortion parameters, but this probably would not help their predictive power. Note (Table 1) the rather large changes in the rotational constants between the different $v_t$ states, as already mentioned in the Introduction.

The fact that the asymmetric rotor model was not very successful in fitting the theoretical levels confirms that it is probably not a good approach for analyzing the experimental spectrum. As an alternative, we adopted a simple model which involves scaling of the theoretical results in order to extend assignment of the spectrum.

## 3. Assignment of the spectra

Spectra were recorded at the University of Calgary using a pulsed supersonic slit jet expansion probed by tunable infrared diode laser ($v_1$ region) or optical parametric oscillator ($v_1 + v_3$ region) sources, as described previously [27,28,38]. Some line positions may differ slightly from Refs. [27,30] because of re-calibration. As an example, Fig. 1 shows corresponding parts of the $v_1$ ($\approx 2063$ cm$^{-1}$) and $v_1 + v_3$ ($\approx 2919$ cm$^{-1}$) band regions for the normal isotopologue of OCS, with labels showing various $R(2)$ transitions of He$_2$-OCS. Other lines, not due to He$_2$-OCS, are labeled with single digits denoting $N$, the number of He atoms in the He$_N$-OCS cluster responsible for the transition. The $v_1$ spectrum was taken with conditions of high backing pressure resulting in the presence of larger clusters up to $N = 8$, whereas the $v_1 + v_3$ spectrum, taken with lower pressure, shows only $N = 0, 1, 2,$ and perhaps 3.

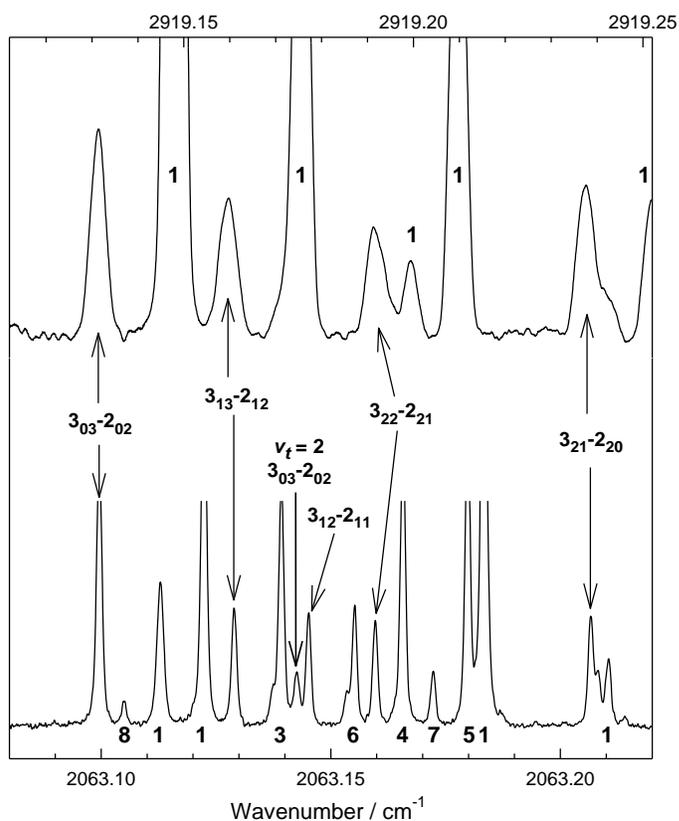

Figure 1. Analogous sections of spectra in the $\nu_1$ and $\nu_1 + \nu_3$ regions with labeled $R(2)$ transitions of He$_2$-OCS. Lines not due to He$_2$-OCS are labeled with single digits denoting $N$, the size of the He$_N$-OCS cluster responsible. The $3_{12} \leftarrow 2_{11}$ and $v_t = 0$, $3_{03} \leftarrow 2_{02}$ transitions for $\nu_1 + \nu_3$ are obscured by a He-OCS transition. Lines in the $\nu_1 + \nu_3$ spectrum are wider due to the higher wavenumber region (greater residual Doppler broadening) and other experimental factors.

In connection with Fig. 1, we note that it is not possible to get anything close to a pure spectrum of He$_2$-OCS. The strength of He$_2$-OCS relative to OCS and He-OCS can be increased by increasing the jet backing pressure and diluting OCS:He gas mixture. But this quickly causes larger He$_N$-OCS clusters to appear, as in the lower trace of Fig. 1.

The strongest and simplest transitions of He$_2$-OCS are those with $v_t = 0$, $K_a = 0$. Many of these were assigned previously [6, 27, 30], and here the list is extended to higher $J$-values and to the OCS $\nu_1 + \nu_3$ combination band (see the top of Table 2). By combining these infrared measurements with highly accurate ground state microwave data [7], it is possible to construct an entirely experimental energy level scheme for $K_a = 0$ up to $J = 6$ in the ground, $\nu_1$, and $\nu_1 + \nu_3$ states (this is done in Sec. 4, below). Comparing these ground state experimental levels with theory [30], we find that experiment is higher than theory by a systematic scaling factor ranging from about 1.0097 to 1.0079 for $J = 1$ to 5. This general result is not surprising, since Paesani and Whaley [21] similarly found that their potential (as used in [30]) resulted in underestimation of the rotational energies of He-OCS. The implication is that He-OCS and He$_2$-OCS are actually slightly more closely bound than given by this potential function (i.e. the He atoms are located closer to the OCS). Though the difference between theory and experiment is small ($\approx$1%), it can still have a crucial effect in deciding between assignments of nearby transitions. Two recent theoretical intermolecular potentials [22, 23] give better agreement with experiment for He-OCS, and presumably would also work well for He$_2$-OCS.

What about the experimental $\nu_1$ excited state $K_a = 0$ levels? After subtracting the band origin, these energies are very similar to the ground state, but slightly reduced (the observed factor is about 0.996), as expected due to the usual effects of anharmonicity. The magnitude of this reduction is somewhat underestimated by theory [30], which predicts about 0.9985. For the $v_1 + v_3$ excited state, the $K_a = 0$ energies are reduced still further (≈0.994) compared to the ground state, but there is no theory in this case. We also have information on the isotopologues He$_2$-O$^{13}$CS and He$_2$-OC$^{34}$S [7, 27], whose $K_a = 0$ energies are found to be reduced by factors of about 0.999 and 0.983, respectively, compared to the normal isotopologue. The very small effect of $^{13}$C substitution on the rotational energies is of course explained by the fact that the C atom lies close to the center of mass of He$_2$-OCS.

The scaling factors mentioned above (which differ only slightly from unity) were useful in extending the He$_2$-OCS assignments to weaker and more difficult transition involving $(v_t, K_a)$ = (0, 2), (1, 1), (1, 3), and (2, 0). It was found that good line position predictions were obtained by scaling the theoretical levels [30], thus avoiding an asymmetric rotor Hamiltonian, whose utility in this case is dubious. Verification of the assignments could then be made using ground state combination differences between the $v_1$ and $v_1 + v_3$ bands, as mentioned above.

Previous assignments for He$_2$-OCS [6, 27, 30] were limited to transitions with $\Delta K_a = 0$, but here we were also able to assign a number of transitions with $K_a = 2 \leftrightarrow 0$ within the $v_t = 0$ state. These $\Delta K_a = 2$ transitions are weaker, but still allowed for an asymmetric rotor (as long as it deviates from the prolate symmetric top limit). Similarly, transitions with $K_a = 3 \leftrightarrow 1$ should be possible within the $v_t = 1$ state, but they are predicted to be even weaker and were not observed.

Table 2. Observed transitions of He$_2$-OCS in the $v_t = 0$ state (in cm$^{-1}$)

| $J\,K_a\,K_c$ ↔ $J\,K_a\,K_c$ | | $\nu_1$ ← | $\nu_1 + \nu_3$ ← | $\nu_1$ → | $\nu_1 + \nu_3$ → |
|---|---|---|---|---|---|
| 1 0 1 | 0 0 0 | 2062.677 | 2918.711 | 2062.174 | 2918.208* |
| 2 0 2 | 1 0 1 | 2062.901 | 2918.935 | 2061.946* | 2917.981* |
| 3 0 3 | 2 0 2 | 2063.100 | 2919.131 | 2061.744 | 2917.777 |
| 4 0 4 | 3 0 3 | 2063.287 | 2919.317 | 2061.551 | 2917.582 |
| 5 0 5 | 4 0 4 | 2063.468* | 2919.495 | 2061.363 | |
| 6 0 6 | 5 0 5 | 2063.643 | 2919.665 | 2061.181 | 2917.209 |
| 2 2 0 | 2 2 1 | 2062.440* | 2918.473 | 2062.403 | 2918.437 |
| 3 2 1 | 3 2 2 | 2062.484 | | 2062.352 | 2918.384 |
| 3 2 1 | 2 2 0 | 2063.207 | 2919.238 | 2061.633* | 2917.666 |
| 4 2 2 | 3 2 1 | 2063.433 | | 2061.397 | 2917.428 |
| 5 2 3 | 4 2 2 | 2063.641 | 2919.665 | 2061.181 | 2917.209 |
| 3 2 2 | 2 2 1 | 2063.160 | 2919.190 | 2061.680 | 2917.714 |
| 4 2 3 | 3 2 2 | 2063.369 | | 2061.465 | 2917.497* |
| 5 2 4 | 4 2 3 | 2063.569 | | 2061.258 | |
| 2 2 1 | 2 0 2 | 2062.711 | 2918.745 | 2062.133* | 2918.166 |
| 3 2 2 | 3 0 3 | 2062.770 | 2918.802 | 2062.069 | 2918.101* |
| 4 2 3 | 4 0 4 | 2062.849 | | 2061.982* | 2918.012* |
| 5 2 4 | 5 0 5 | 2062.949* | 2918.973* | 2061.877* | 2917.903* |
| 2 2 0 | 1 0 1 | 2063.208 | | | |
| 3 2 1 | 2 0 2 | | 2919.548* | 2061.326* | |
| 4 2 2 | 3 0 3 | 2063.851* | | | |

Note: Transitions labeled with an asterisk are less certain due to blending or weakness.

Table 3. Observed transitions of He$_2$-OCS in the $v_t = 1$ and 2 states (in cm$^{-1}$)

| $v_t$ | $J\,K_a\,K_c$ | $\leftrightarrow$ | $J\,K_a\,K_c$ | $\nu_1$ $\leftarrow$ | $\nu_1 + \nu_3$ $\leftarrow$ | $\nu_1$ $\rightarrow$ | $\nu_1 + \nu_3$ $\rightarrow$ |
|---|---|---|---|---|---|---|---|
| 1 | 1 1 1 | | 1 1 0 | 2062.411* | 2918.444 | 2062.428 | 2918.462 |
| 1 | 2 1 2 | | 2 1 1 | 2062.396 | 2918.428 | 2062.440* | 2918.473 |
| 1 | 3 1 3 | | 3 1 2 | | 2918.406* | 2062.455* | 2918.485 |
| 1 | 2 1 2 | | 1 1 1 | 2062.894 | 2918.927 | 2061.944* | 2917.976* |
| 1 | 3 1 3 | | 2 1 2 | 2063.129 | 2919.160 | 2061.705 | 2917.738 |
| 1 | 4 1 4 | | 3 1 3 | 2063.360 | 2919.388 | 2061.469 | 2917.499* |
| 1 | 5 1 5 | | 4 1 4 | 2063.581 | 2919.606 | 2061.240 | 2917.269* |
| 1 | 2 1 1 | | 1 1 0 | 2062.908* | | 2061.929 | 2917.963* |
| 1 | 3 1 2 | | 2 1 1 | 2063.145 | 2919.176* | 2061.687 | 2917.720* |
| 1 | 4 1 3 | | 3 1 2 | 2063.376 | 2919.404 | 2061.449 | 2917.479 |
| 1 | 5 1 4 | | 4 1 3 | 2063.599 | 2919.624* | 2061.216* | 2917.244* |
| 1 | 3 3 - | | 3 3 - | 2062.411* | 2918.444* | -- | -- |
| 1 | 4 3 - | | 3 3 - | 2063.386* | 2919.415* | | |
| 1 | 5 3 - | | 4 3 - | 2063.604 | 2919.637* | 2061.208* | |
| 2 | 1 0 1 | | 0 0 0 | 2062.665 | 2918.698 | 2062.162 | 2918.196 |
| 2 | 2 0 2 | | 1 0 1 | | | 2061.915 | 2917.947 |
| 2 | 3 0 3 | | 2 0 2 | 2063.143 | | 2061.678 | |
| 2 | 4 0 4 | | 3 0 3 | 2063.373 | | 2061.441 | |
| 2 | 5 0 5 | | 4 0 4 | | | 2061.196* | |

Note: Transitions labeled with an asterisk are less certain due to blending or weakness. The $K_c$ quantum number is omitted for $K_a = 3$ transitions because asymmetry doubling was not resolved.

The resulting list of assignments for the $v_t = 0$, 1, and 2 states is given in Tables 2 and 3 for the $v_1$ and $v_1 + v_3$ bands of the normal isotopologue of He$_2$-OCS. Analogous lists for the $v_1$ bands of He$_2$-O$^{13}$CS and He$_2$-OC$^{34}$S are given as Supplementary Data. Transitions missing from these tables were obscured by lines of OCS, He-OCS, etc. Unfortunately, no transitions of significant intensity connecting different $v_t$ states were predicted [30], and none were observed. Tables 2 and 3 are arranged with "mirror image" transitions on the same row, e.g. $(J, K_a, K_c) = (2, 0, 2) \leftarrow (1, 0, 1)$ and $(1, 0, 1) \leftarrow (2, 0, 2)$. They contain 129 assigned He$_2$-OCS infrared transitions in the the $v_1$ and $v_1 + v_3$ bands, a significant increase compared to Ref. [30] which had 23 transitions in the $v_1$ band. Note that there are few key re-assignments, notably the $J = 1 \leftarrow 2$ and $2 \leftarrow 3$ transitions for $K_a = 1$ (Table 3), and also some small changes due to re-calibration, as mentioned above.

## 4. He$_2$-OCS Energy Levels

As already noted, purely experimental energies can be derived for $v_t = 0$, $K_a = 0$ levels up to $J = 6$ by combining infrared and microwave data [7]. Thanks to the new $K_a = 2 \leftrightarrow 0$ transitions assigned in Table 2, the $v_t = 0$, $K_a = 2$ levels can now be included in the same scheme. The resulting set of experimental $v_t = 0$ energy levels is compared with theory in Table 4 for the ground vibrational state of the normal isotopologue. Analogous lists are given as Supplementary Data for the $v_1$ and $v_1 + v_3$ excited states, and for the He$_2$-O$^{13}$CS and He$_2$-OC$^{34}$S isotopologues. The accuracy of the experimental energies is not as good as quoted (0.0001 cm$^{-1}$), except for $(J, K_a, K_c) = (1, 0, 1)$, (2, 0, 2), and (3, 0, 3) which are very well determined by microwave data [7]. But on the other hand the accuracy should be better than 0.001 cm$^{-1}$, at least in favorable cases. The ratios given in the last column of Table 4 illustrate the fairly uniform scaling between experiment and theory that was discussed above, and also show that the scaling factor is somewhat larger for $K_a = 2$ than for $K_a = 0$.

Table 4. Theory compared with experimental energies for $v_t = 0$, and quasi-experimental energies for $v_t = 1$, $K_a = 1$, of He$_2$-OCS (in cm$^{-1}$).

| $v_t$ | $J\ K_a\ K_c$ | Experiment | Theory [30] | Exp/Theory |
|---|---|---|---|---|
| 0 | 0 0 0 | 0.0 | 0.0 | |
| 0 | 1 0 1 | 0.25233 | 0.2499 | 1.0097 |
| 0 | 2 0 2 | 0.73080 | 0.7243 | 1.0090 |
| 0 | 3 0 3 | 1.41015 | 1.3988 | 1.0081 |
| 0 | 4 0 4 | 2.2801 | 2.2628 | 1.0076 |
| 0 | 5 0 5 | 3.3350 | 3.3090 | 1.0079 |
| 0 | 6 0 6 | 4.5673 | -- | -- |
| 0 | 2 2 1 | 1.0203 | 1.0001 | 1.0202 |
| 0 | 3 2 2 | 1.7617 | 1.7303 | 1.0181 |
| 0 | 4 2 3 | 2.7153 | 2.6697 | 1.0171 |
| 0 | 5 2 4 | 3.8723 | 3.8140 | 1.0153 |
| 0 | 2 2 0 | 1.0392 | 1.0175 | 1.0213 |
| 0 | 3 2 1 | 1.8283 | 1.7909 | 1.0209 |
| 0 | 4 2 2 | 2.8485 | 2.7884 | 1.0216 |
| 0 | 5 2 3 | 4.0808 | 3.9951 | 1.0215 |
| 1 | 1 1 1 | 0.3152 | 0.3152 | |
| 1 | 2 1 2 | 0.7915 | 0.7898 | 1.0021 |
| 1 | 3 1 3 | 1.5046 | 1.5022 | 1.0016 |
| 1 | 4 1 4 | 2.4519 | 2.4471 | 1.0020 |
| 1 | 5 1 5 | 3.6243 | 3.6136 | 1.0030 |
| 1 | 1 1 0 | 0.3236 | 0.3217 | 1.0060 |
| 1 | 2 1 1 | 0.8137 | 0.8071 | 1.0082 |
| 1 | 3 1 2 | 1.5442 | 1.5306 | 1.0089 |
| 1 | 4 1 3 | 2.5097 | 2.4856 | 1.0097 |
| 1 | 5 1 4 | 3.7034 | 3.6641 | 1.0107 |

It is not possible to derive similar sets of pure experimental energies for the other states observed here, $(v_t, K_a) = (1, 1), (1, 3)$, and $(2, 0)$. There are two problems. First, due to the absence of $K_a \ne 0$ microwave data, there is no way to fix the relative energies of ground state levels of different parity: for example, there is no experimental information giving the interval between the ground state $(J, K_a, K_c) = (1, 1, 0)$ and $(1, 1, 1)$ levels for $v_t = 1$. (See Table 3 of [30] for the level parities; the needed microwave transitions should be observable in the future.) Second, there are no observed transitions connecting different $v_t$ states, nor any observed $K_a = 3 \leftrightarrow 1$ transitions within $v_t = 1$. In order to work around the first problem, we used a simple model to derive a "quasi-experimental" set of energies for $(v_t, K_a) = (1, 1)$ independent of theory by requiring a uniform ratio between the ground and excited state ($\nu_1$ or $\nu_1 + \nu_3$) energies. This ratio was simply a fitting parameter, used to compensate for the fact that we do not know the relative energy of the two parity sets. Results are given in the lower part of Table 4 for the ground state, and in the Supplementary Data for the excited states. Optimum values for the ratios were 0.9961 for $\nu_1$ and 0.9936 for $\nu_1 + \nu_3$. This approximation (the same uniform ratio between the excited and ground state for all levels) is doubtless imperfect, but the resulting energies actually fit the observed spectra quite well. In Table 4, the energy of the lowest $(J, K_a, K_c) = (1, 1, 1)$ level is arbitrarily fixed at its calculated value of 0.3152 cm$^{-1}$ relative to the forbidden $(0, 0, 0)$ level of $v_t = 1$ (itself calculated to be 0.4296 cm$^{-1}$ above $v_t = 0$). The ratios shown in the last column of Table 4 suggest that theory predicts the energies better for the lower $K_a = 1$ asymmetry components ($(1, 1, 1), (2, 1, 2)$ etc.) than for the higher components ($(1, 1, 0)$, etc.). Similar "quasi-experimental" results for $(v_t, K_a) = (1, 1)$ levels of He$_2$-O$^{13}$CS and He$_2$-OC$^{34}$S are given in Supplementary Data.

The assignments for $(v_t, K_a) = (2, 0)$ and $(1, 3)$ transitions are too fragmentary to support experimental energy lists like Table 4. However, it is worth noting that the $(v_t, K_a) = (2, 0)$ observations are consistent with a factor of 1.008 between experimental and theoretical energies, very similar to the factor observed for $(v_t, K_a) = (0, 0)$ (Table 4). Similarly, the $(v_t, K_a) = (1, 3)$ observations are consistent with a factor of about 1.006. The effective $\nu_1$ band origins for $(v_t, K_a) = (2, 0)$ and $(1, 3)$ are about 2062.415 and 2062.426 cm$^{-1}$, which may be compared with values of 2062.426 cm$^{-1}$ for $(v_t, K_a) = (0, 0)$ and 2062.421 cm$^{-1}$ for $(v_t, K_a) = (1, 1)$.

Figure 2 shows the observed experimental rotational levels of He$_2$-OCS for the ground vibrational state, with theoretical levels [30] indicated by dotted lines. The energy of $v_t = 1$ relative to $v_t = 0$ is not experimentally known, so the lowest level is fixed to its theoretical value. This figure graphically illustrates the small but crucial differences between experiment and theory. It also emphasizes the slightly surprising fact (evident in the spectra) that asymmetry splittings are larger for $K_a = 2$ than for $K_a = 1$ (for $J > 2$), a reflection of the differing rotational constants of $v_t = 0$ and 1.

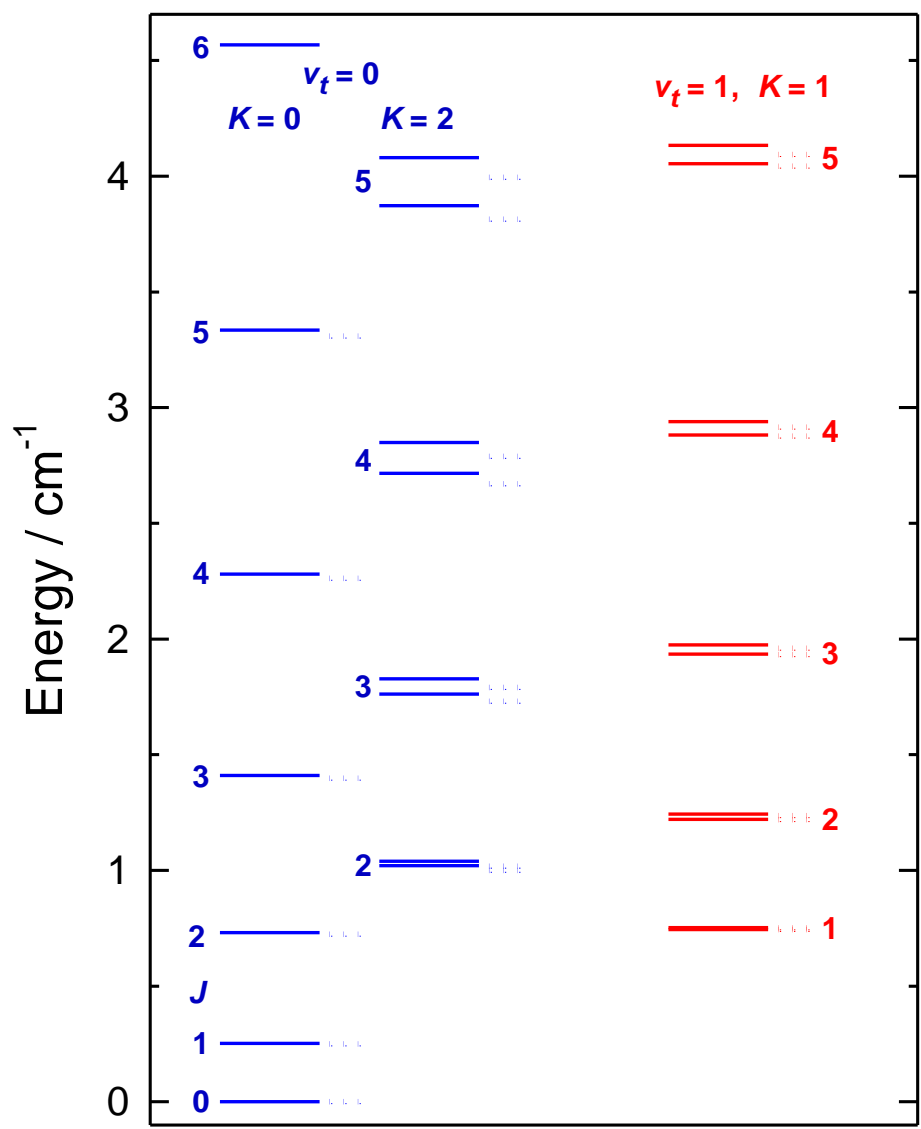

Figure 2. Experimental (solid lines) and theoretical (dotted lines) energy levels for the ground state of He$_2$-OCS. The relative energy of torsional states $v_t = 0$ and 1 is not known experimentally, so the lowest experimental $v_t = 1$ level is fixed at its theoretical value.

## 5. Discussion and conclusions

To date, He$_2$-OCS transitions within the $(v_t, K_a) = (0, 0)$ stack are the only ones that have been observed in the microwave region. But there is no reason why pure rotational transitions within other stacks, namely (0, 2), (1, 1) and (2, 0), should not also be detected. In Table 5, we list some predicted microwave transitions, based on our experimental ground state energy levels. Though it is difficult to know how reliable these predictions are, we are optimistic that they should be better than 20, 40, and 80 MHz for $v_t = 0$, 1, and 2, respectively.

Table 5. Some predicted He$_2$-OCS microwave transitions (in MHz).

| $v_t$ | $J' K_a' K_c'$ | $J'' K_a'' K_c''$ | Frequency |
|---|---|---|---|
| 0 | 3 2 1 | 3 2 2 | 1997 |
| 0 | 4 2 2 | 4 2 3 | 3993 |
| 0 | 3 2 2 | 2 2 1 | 22227 |
| 0 | 3 2 1 | 2 2 0 | 23657 |
| 0 | 4 2 3 | 3 2 2 | 28588 |
| 0 | 4 2 2 | 3 2 1 | 30585 |
| 0 | 2 2 1 | 2 0 2 | 8679 |
| 0 | 3 2 2 | 3 0 3 | 10359 |
| 0 | 4 2 3 | 4 0 4 | 13047 |
| 1 | 2 1 2 | 1 1 1 | 14264 |
| 1 | 2 1 1 | 1 1 0 | 14691 |
| 1 | 3 1 3 | 2 1 2 | 21380 |
| 1 | 3 1 2 | 2 1 1 | 21899 |
| 1 | 4 1 4 | 3 1 3 | 28397 |
| 1 | 4 1 3 | 3 1 2 | 28943 |
| 2 | 1 0 1 | 0 0 0 | 7598 |
| 2 | 2 0 2 | 1 0 1 | 14964 |
| 2 | 3 0 3 | 2 0 2 | 21979 |

Vibrational shifts for He$_N$-OCS clusters have already been extensively studied, but only in the OCS $\nu_1$ region [6, 9, 27], where it is well known from previous work that the shift for He$_2$-OCS is very close to twice that for He-OCS. The new information from the present work is the He$_2$-OCS shift for the $\nu_1 + \nu_3$ combination band, which is +0.3708 cm$^{-1}$, which is about 2.11 times the observed He-OCS shift [28].

It would be especially interesting to detect and assign transitions between different $v_t$ states, as was done (though only tentatively) for the similar He$_2$-N$_2$O complex [36]. To acquire significant intensity, such transitions rely on the mixing caused by a close coincidence between levels with equal $J$ and different $v_t$ values. The best coincidence candidate in the current set of levels seems to be $(J, K_a, K_c) = (5, 1, 4)$ and $(5, 2, 3)$ (the upper asymmetry components, see Fig. 2). But there is considerable uncertainty in their relative energy.

In conclusion, we have extended the analysis of the spectrum of He$_2$-OCS by assigning many new transitions for higher levels of $J$, $K_a$, and $v_t$ (the low frequency torsional vibration). The analysis relied upon judicious scaling of calculated energy levels [30], rather than fitting spectra using a theoretical Hamiltonian, and assignments were validated by comparing combination differences for two He$_2$-OCS infrared bands ($\nu_1$ and $\nu_1 + \nu_3$) having a common ground state. The present study confirms that the calculated energies of Ref. [30] are rather good, especially after allowing for simple scaling ($\approx 1.008$) -- indeed rather better than one might have thought simply on the basis of the comparisons given there [30].

**Acknowledgments**

Financial support from the Natural Sciences and Engineering Research Council of Canada, and the Canadian Space agency, is gratefully acknowledged.

**Appendix A. Supplementary material**

Supplementary data associated with this article can be found below.

**Supplementary Data for:**

# The He$_2$ - OCS complex: comparison between theory and experiment

by

J. Norooz Oliaee,[1] N. Moazzen-Ahmadi,[1] A.R.W. McKellar,[2] Xiao-Gang Wang,[3]
and Tucker Carrington, Jr.[3]

[1] *Department of Physics and Astronomy, University of Calgary, 2500 University Drive North West, Calgary, Alberta T2N 1N4, Canada.*

[2] *National Research Council of Canada, Ottawa, Ontario K1A 0R6, Canada.*

[3] *Chemistry Department, Queen's University, Kingston, Ontario K7L 3N6, Canada.*

Table A-1. Theoretical calculated energy levels for $He_2 - OCS$ (in $cm^{-1}$).

| J | $K_a$ | $K_c$ | $v_t = 0$ | $v_t = 1$ | $v_t = 2$ |
|---|---|---|---|---|---|
| 0 | 0 | 0 | 0.0000 | 0.4296 | 1.5845 |
| 1 | 0 | 1 | 0.2499 | 0.6683 | 1.8358 |
| 1 | 1 | 1 | 0.2932 | 0.7448 | 1.8866 |
| 1 | 1 | 0 | 0.3368 | 0.7513 | 1.9216 |
| 2 | 0 | 2 | 0.7243 | 1.1445 | 2.3307 |
| 2 | 1 | 2 | 0.7468 | 1.2194 | 2.3542 |
| 2 | 1 | 1 | 0.8713 | 1.2367 | 2.4626 |
| 2 | 2 | 1 | 1.0001 | 1.4682 | 2.6020 |
| 2 | 2 | 0 | 1.0175 | 1.4690 | 2.6172 |
| 3 | 0 | 3 | 1.3988 | 1.8569 | 3.0576 |
| 3 | 1 | 3 | 1.4085 | 1.9318 | 3.0603 |
| 3 | 1 | 2 | 1.6366 | 1.9602 | 3.2807 |
| 3 | 2 | 2 | 1.7303 | 2.1960 | 3.3568 |
| 3 | 2 | 1 | 1.7909 | 2.1983 | 3.4302 |
| 3 | 3 | 1 | 2.0474 | 2.5781 | 3.6904 |
| 3 | 3 | 0 | 2.0502 | 2.5772 | 3.6961 |
| 4 | 0 | 4 | 2.2628 | 2.8024 | 4.0175 |
| 4 | 1 | 4 | 2.2665 | 2.8767 | 3.9920 |
| 4 | 1 | 3 | 2.6080 | 2.9152 | 4.3675 |
| 4 | 2 | 3 | 2.6697 | 3.1587 | 4.3776 |
| 4 | 2 | 2 | 2.7884 | 3.1616 | 4.5679 |
| 4 | 3 | 2 | 3.0144 | 3.5515 | 4.7229 |
| 4 | 3 | 1 | 3.0248 | 3.5487 | 4.7660 |
| 4 | 4 | 1 | 3.4305 | 4.0771 | 5.1554 |
| 4 | 4 | 0 | 3.4272 | 4.0520 | 5.1573 |
| 5 | 0 | 5 | 3.3090 | 3.9743 | 5.2213 |
| 5 | 1 | 5 | 3.3102 | 4.0432 | 5.2171 |
| 5 | 1 | 4 | 3.7749 | 4.0937 | 5.6832 |
| 5 | 2 | 4 | 3.8140 | 4.3362 | 5.6487 |
| 5 | 2 | 3 | 3.9951 | 4.3354 | 6.0395 |
| 5 | 3 | 3 | 4.1905 | 4.7438 | 6.0539 |
| 5 | 3 | 2 | 4.2109 | 4.7408 | 6.2063 |
| 5 | 4 | 2 | 4.6129 | 5.2912 | 6.4543 |
| 5 | 4 | 1 | 4.5977 | 5.2439 | 6.4751 |
| 5 | 5 | 1 | 5.1092 | 5.9045 | 6.9872 |
| 5 | 5 | 0 | 5.1468 | 5.8627 | 6.9797 |

**Notes:** Taken from: X.-G. Wang, T. Carrington, Jr., Can. J. Chem. 88 (2010) 779.

Based on the He – OCS potential surface of: F. Paesani, K.B. Whaley, J. Chem. Phys. 121 (2004) 4180.

Some energies here may differ slightly (0.0001 $cm^{-1}$) from those in the original paper.

Nuclear spin symmetry forbids levels with ($v_t$, $K_a$) = (even, odd) and (odd, even) for $^4He_2$-OCS, but they can still be calculated, as given here.

Table A-2. Asymmetric rotor fits to He$_2$ – OCS theoretical energy levels (in cm$^{-1}$).

|  | $v_t = 0$ | $v_t = 1$ | $v_t = 2$ |
|---|---|---|---|
| T$_0$ | 0.0000 | 0.4296 | 1.5845 |
| A | 0.18907(38) | 0.20122(36) | 0.19254(39) |
| B | 0.14893(44) | 0.1416(14) | 0.14230(51) |
| C | 0.10207(44) | 0.1164(14) | 0.10781(41) |
| $10^3 \times \Delta_K$ | 0.494(49) | 0.756(47) | -0.807(49) |
| $10^3 \times \Delta_{JK}$ | -0.475(55) | -0.405(55) | 1.112(62) |
| $10^3 \times \Delta_J$ | 0.315(13) | 0.033(14) | -0.304(16) |
| $10^3 \times \delta_K$ | 0.652(72) | -0.009(657) | -0.153(41) |
| $10^3 \times \delta_J$ | 0.110(11) | 0.030(13) | -0.048(11) |
| rms error | 0.0125 | 0.0150 | 0.0091 |

**Notes:** All levels with $J = 1$ to 4 from Table A-1 were fitted (including "forbidden" ones). The rms errors refer to the root mean square deviations for all levels with $J = 1$ to 5, so in part they measures the success (or otherwise) of the fit to predict the $J = 5$ levels.

Table A-3. Observed transitions of He$_2$-O$^{13}$CS and He$_2$-OC$^{34}$S for the $v_t = 0$ state in the OCS $v_1$ band (in cm$^{-1}$).

| $J\,K_a\,K_c$ ↔ $J\,K_a\,K_c$ | | He$_2$-O$^{13}$CS | | He$_2$-OC$^{34}$S | |
|---|---|---|---|---|---|
| | | ← | → | ← | → |
| 1 0 1 | 0 0 0 | 2009.697 | 2009.195 | 2061.916 | 2061.423* |
| 2 0 2 | 1 0 1 | 2009.922 | 2008.968 | 2062.137* | 2061.198 |
| 3 0 3 | 2 0 2 | 2010.120 | 2008.765 | 2062.334 | 2060.998* |
| 4 0 4 | 3 0 3 | 2010.308 | 2008.573 | 2062.518* | 2060.808 |
| 5 0 5 | 4 0 4 | 2010.488 | 2008.385 | 2062.697 | 2060.621* |
| 6 0 6 | 5 0 5 | 2010.662 | 2008.203* | 2062.869 | 2060.442 |
| 2 2 0 | 2 2 1 | 2009.461 | 2009.424 | *2061.683* | 2061.649 |
| 3 2 1 | 3 2 2 | 2009.504 | 2009.374 | | 2061.603 |
| 3 2 1 | 2 2 0 | 2010.227 | 2008.655 | 2062.431* | 2060.896* |
| 3 2 2 | 2 2 1 | 2010.179 | 2008.702 | 2062.388 | 2060.941 |
| 4 2 2 | 3 2 1 | 2010.453 | | 2062.657* | 2060.663* |
| 4 2 3 | 3 2 2 | 2010.389 | 2008.486 | 2062.595 | 2060.727* |
| 5 2 3 | 4 2 2 | 2010.659 | 2008.203* | 2062.861 | 2060.448 |
| 5 2 4 | 4 2 3 | 2010.590 | 2008.278 | 2062.794* | |
| 2 2 1 | 2 0 2 | 2009.734* | 2009.154* | 2061.9633 | 2061.370* |
| 3 2 2 | 3 0 3 | 2009.792 | 2009.091 | 2062.0165 | 2061.310* |
| 4 2 3 | 4 0 4 | | 2009.002* | | |
| 5 2 4 | 5 0 5 | | 2008.896* | | 2062.184* |
| 2 2 0 | 1 0 1 | | | | 2062.450* |
| 3 2 1 | 2 0 2 | | 2008.349* | | 2062.745 |

Note: Transitions labeled with an asterisk are less certain due to blending or weakness.

Table A-4. Observed transitions of He$_2$-O$^{13}$CS and He$_2$-OC$^{34}$S for the $v_t = 1$ state in the OCS $v_1$ band (in cm$^{-1}$)

| $J\,K_a\,K_c$ ↔ $J\,K_a\,K_c$ | | He$_2$-O$^{13}$CS $v_1$ | | He$_2$-OC$^{34}$S $v_1$ | |
|---|---|---|---|---|---|
| | | ← | → | ← | → |
| 1 1 1 | 1 1 0 | 2009.433* | 2009.449 | 2061.657 | 2061.672 |
| 2 1 2 | 2 1 1 | 2009.417 | 2009.461 | 2061.642 | 2061.683 |
| 3 1 3 | 3 1 2 | | 2009.472 | | |
| 2 1 2 | 1 1 1 | 2009.915 | 2008.966 | 2062.129 | |
| 3 1 3 | 2 1 2 | 2010.149 | 2008.727 | 2062.359 | 2060.964* |
| 4 1 4 | 3 1 3 | 2010.380 | 2008.491 | 2062.586 | 2060.731 |
| 5 1 5 | 4 1 4 | | 2008.263 | 2062.804 | |
| 2 1 1 | 1 1 0 | 2009.928* | 2008.951 | | 2061.183 |
| 3 1 2 | 2 1 1 | 2010.165 | 2008.709 | 2062.375 | 2060.946 |
| 4 1 3 | 3 1 2 | 2010.396 | 2008.471 | | 2060.712 |
| 5 1 4 | 4 1 3 | 2010.620 | 2008.240 | 2062.828* | |
| 3 3 | 3 3 | 2009.433 | | 2061.657 | |
| 4 3 | 3 3 | | | | |
| 5 3 | 4 3 | 2010.624 | 2008.227 | | |

Note: Transitions labeled with an asterisk are less certain due to blending or weakness.

Table A-5. Observed transitions of He$_2$-O$^{13}$CS and He$_2$-OC$^{34}$S for the $v_t = 2$ state in the OCS $v_1$ band (in cm$^{-1}$)

| $J\,K_a\,K_c$ ↔ $J\,K_a\,K_c$ | | He$_2$-O$^{13}$CS $v_1$ | | He$_2$-OC$^{34}$S $v_1$ | |
|---|---|---|---|---|---|
| | | ← | → | ← | → |
| 1 0 1 | 0 0 0 | 2009.686* | 2009.184 | 2061.904 | 2061.411* |
| 2 0 2 | 1 0 1 | | 2008.936 | 2062.150* | |
| 3 0 3 | 2 0 2 | 2010.163 | 2008.700 | 2062.369 | 2060.936* |
| 4 0 4 | 3 0 3 | | 2008.463 | | 2060.704* |
| 5 0 5 | 4 0 4 | 2010.629* | 2008.219 | 2062.827* | |

Note: Transitions labeled with an asterisk are less certain due to blending or weakness.

Table A-6. Comparison of observed and unscaled theoretical transitions of He$_2$-OCS in the $v_t = 0$ state (in cm$^{-1}$)

| $J\,K_a\,K_c$ ↔ $J\,K_a\,K_c$ | | $v_1$ Obs ← | $v_1$ Calc ← | O-C ← | $v_1$ Obs → | $v_1$ Calc → | O-C → |
|---|---|---|---|---|---|---|---|
| 1 0 1 | 0 0 0 | 2062.677 | 2062.676 | 0.001 | 2062.174 | 2062.176 | -0.002 |
| 2 0 2 | 1 0 1 | 2062.901 | 2062.900 | 0.001 | 2061.946* | 2061.951 | -0.005 |
| 3 0 3 | 2 0 2 | 2063.100 | 2063.099 | 0.001 | 2061.744 | 2061.751 | -0.007 |
| 4 0 4 | 3 0 3 | 2063.287 | 2063.287 | 0.000 | 2061.551 | 2061.560 | -0.009 |
| 5 0 5 | 4 0 4 | 2063.468* | 2063.477 | -0.009 | 2061.363 | 2061.377 | -0.014 |
| 6 0 6 | 5 0 5 | 2063.643 | | | 2061.181 | | |
| 2 2 0 | 2 2 1 | 2062.440* | 2062.441 | -0.001 | 2062.403 | 2062.406 | -0.003 |
| 3 2 1 | 3 2 2 | 2062.484 | 2062.483 | 0.001 | 2062.352 | 2062.362 | -0.010 |
| 3 2 1 | 2 2 0 | 2063.207 | 2063.195 | 0.012 | 2061.633* | 2061.650 | -0.017 |
| 4 2 2 | 3 2 1 | 2063.433 | 2063.418 | 0.015 | 2061.397 | 2061.425 | -0.028 |
| 5 2 3 | 4 2 2 | 2063.641 | 2063.625 | 0.016 | 2061.181 | 2061.214 | -0.033 |
| 3 2 2 | 2 2 1 | 2063.160 | 2063.152 | 0.008 | 2061.680 | 2061.693 | -0.013 |
| 4 2 3 | 3 2 2 | 2063.369 | 2063.361 | 0.008 | 2061.465 | 2061.483 | -0.018 |
| 5 2 4 | 4 2 3 | 2063.569 | 2063.564 | 0.005 | 2061.258 | 2061.277 | -0.019 |
| 2 2 1 | 2 0 2 | 2062.711 | 2062.700 | 0.011 | 2062.133* | 2062.150 | -0.017 |
| 3 2 2 | 3 0 3 | 2062.770 | 2062.754 | 0.016 | 2062.069 | 2062.093 | -0.024 |
| 4 2 3 | 4 0 4 | 2062.849 | 2062.829 | 0.020 | 2061.982* | 2062.017 | -0.035 |
| 5 2 4 | 5 0 5 | 2062.949* | 2062.925 | 0.024 | 2061.877* | 2061.917 | -0.040 |
| 2 2 0 | 1 0 1 | 2063.208 | 2063.191 | 0.017 | | 2061.659 | |
| 3 2 1 | 2 0 2 | | 2063.489 | | 2061.326* | 2061.359 | -0.033 |
| 4 2 2 | 3 0 3 | 2063.851* | 2063.810 | 0.041 | | 2061.035 | -0.002 |

Note: Transitions labeled with an asterisk are less certain due to blending or weakness.

Table A-7. Comparison of observed and unscaled theoretical transitions of He$_2$-OCS in the $v_t = 1$ state (in cm$^{-1}$)

| $J\,K_a\,K_c$ ↔ $J\,K_a\,K_c$ | | $v_1$ Obs ← | $v_1$ Calc ← | O-C ← | $v_1$ Obs → | $v_1$ Calc → | O-C → |
|---|---|---|---|---|---|---|---|
| 1 1 1 | 1 1 0 | 2062.411* | 2062.416 | -0.005 | 2062.428 | 2062.428 | 0.000 |
| 2 1 2 | 2 1 1 | 2062.396 | 2062.404 | -0.008 | 2062.440* | 2062.438 | 0.002 |
| 3 1 3 | 3 1 2 | | | | 2062.455* | 2062.448 | 0.007 |
| 2 1 2 | 1 1 1 | 2062.894 | 2062.896 | -0.002 | 2061.944* | 2061.947 | -0.003 |
| 3 1 3 | 2 1 2 | 2063.129 | 2063.133 | -0.004 | 2061.705 | 2061.709 | -0.004 |
| 4 1 4 | 3 1 3 | 2063.360 | 2063.365 | -0.005 | 2061.469 | 2061.476 | -0.007 |
| 5 1 5 | 4 1 4 | 2063.581 | 2063.585 | -0.004 | 2061.240 | 2061.254 | -0.014 |
| 2 1 1 | 1 1 0 | 2062.908* | 2062.907 | 0.001 | 2061.929 | 2061.936 | -0.007 |
| 3 1 2 | 2 1 1 | 2063.145 | 2063.143 | 0.002 | 2061.687 | 2061.698 | -0.011 |
| 4 1 3 | 3 1 2 | 2063.376 | 2063.373 | 0.003 | 2061.449 | 2061.465 | -0.016 |
| 5 1 4 | 4 1 3 | 2063.599 | 2063.595 | 0.004 | 2061.216* | 2061.240 | -0.024 |
| 3 3 | 3 3 | 2062.411* | 2062.415 | -0.004 | -- | -- | |
| 4 3 | 3 3 | 2063.386* | 2063.387 | -0.001 | | | |
| 5 3 | 4 3 | 2063.604 | 2063.605 | -0.001 | 2061.208* | 2061.222 | -0.014 |

Note: Transitions labeled with an asterisk are less certain due to blending or weakness. The $K_c$ quantum number is not given for $K_a = 3$ transitions, since asymmetry doubling was not resolved.

Table A-8. Comparison of observed and unscaled theoretical transitions of He$_2$-OCS in the $v_t = 2$ state (in cm$^{-1}$)

| $J\,K_a\,K_c$ ↔ $J\,K_a\,K_c$ | | $v_1$ Obs ← | $v_1$ Calc ← | O-C ← | $v_1$ Obs → | $v_1$ Calc → | O-C → |
|---|---|---|---|---|---|---|---|
| 1 0 1 | 0 0 0 | 2062.665 | 2062.669 | -0.004 | 2062.162 | 2062.166 | -0.004 |
| 2 0 2 | 1 0 1 | | | | 2061.915 | 2062.912 | 0.003 |
| 3 0 3 | 2 0 2 | 2063.143 | 2063.143 | 0.000 | 2061.678 | 2061.690 | -0.012 |
| 4 0 4 | 3 0 3 | 2063.373 | 2063.374 | -0.001 | 2061.441 | 2061.456 | -0.015 |
| 5 0 5 | 4 0 4 | | | | 2061.196* | 2061.211 | -0.015 |

Note: Transitions labeled with an asterisk are less certain due to blending or weakness.

Table A-9. Comparison of experimental and theoretical energies for $v_t = 0$ levels of excited state (OCS $\nu_1$ or $\nu_1 + \nu_3$) $He_2$-OCS (in cm$^{-1}$).

| $J\ K_a\ K_c$ | Experiment $\nu_1$ | Theory $\nu_1$ | Exp/Theory $\nu_1$ | Experiment $\nu_1 + \nu_3$ |
|---|---|---|---|---|
| 0 0 0 | 2062.4258 | | | 2918.4757 |
| 1 0 1 | 0.2509 | 0.2496 | 1.0053 | 0.2506 |
| 2 0 2 | 0.7278 | 0.7233 | 1.0062 | 0.7263 |
| 3 0 3 | 1.4047 | 1.3969 | 1.0056 | 1.4013 |
| 4 0 4 | 2.2717 | 2.2596 | 1.0054 | 2.2663 |
| 5 0 5 | 3.3223 | 3.3041 | 1.0055 | 3.3149 |
| 6 0 6 | 4.5526 | -- | -- | 4.540* |
| 2 2 1 | 1.0160 | 0.9973 | 1.0188 | 1.0150 |
| 3 2 2 | 1.7542 | 1.7265 | 1.0160 | 1.7512 |
| 4 2 3 | 2.7044 | 2.6647 | 1.0149 | |
| 5 2 4 | 3.8585 | 3.8075 | 1.0134 | 3.847* |
| 2 2 0 | 1.0347 | 1.0147 | 1.0197 | 1.0331 |
| 3 2 1 | 1.8200 | 1.7868 | 1.0186 | 1.8159 |
| 4 2 2 | 2.8357 | 2.7823 | 1.0192 | 2.8287 |
| 5 2 3 | 4.0637 | 3.9866 | 1.0193 | 4.0527 |

Note: Experimental vibrational origins are given for $(J\ K_a\ K_c) = (0\ 0\ 0)$, and the remaining energies are given relative to these origins.

Levels marked with * are less certain.

Table A-10. Experimental energies for $v_t = 0$ levels of ground and $v_1$ excited state He$_2$-O$^{13}$CS and He$_2$-OC$^{34}$S (in cm$^{-1}$).

| $J\ K_a\ K_c$ | He$_2$-O$^{13}$CS | | He$_2$-OC$^{34}$S | |
|---|---|---|---|---|
| | Ground | $v_1$ | Ground | $v_1$ |
| 0 0 0 | 0.0000 | 2009.4468 | 0.0000 | 2061.6693 |
| 1 0 1 | 0.2521 | 0.2506 | 0.2466 | 0.2460 |
| 2 0 2 | 0.7302 | 0.7275 | 0.7169 | 0.7144 |
| 3 0 3 | 1.4092 | 1.4038 | 1.3857 | 1.3810 |
| 4 0 4 | 2.2782 | 2.2705 | 2.2426 | 2.2349 |
| 5 0 5 | 3.3328 | 3.3200 | 3.2834 | 3.2705 |
| 6 0 6 | 4.5633 | 4.5477 | 4.4976 | 4.4827 |
| 2 2 1 | 1.0215 | 1.0167 | 1.0139 | 1.0110 |
| 3 2 2 | 1.7613 | 1.7542 | 1.7398 | 1.7328 |
| 4 2 3 | 2.7146 | 2.7033 | 2.6750 | 2.6658 |
| 5 2 4 | 3.8719 | 3.8579 | 3.8097 | 3.7988 |
| 2 2 0 | 1.0395 | 1.0354 | 1.0309 | 1.0271 |
| 3 2 1 | 1.8273 | 1.8190 | 1.7998 | 1.7930 |
| 4 2 2 | | 2.8332 | 2.7993 | 2.7871 |
| 5 2 3 | 4.0766 | | 4.0087 | 3.9907 |

Note: Experimental vibrational origins are given for $(J\ K_a\ K_c) = (0\ 0\ 0)$, and the remaining $v_1$ state energies are given relative to these origins.

Table A-11. Quasi-experimental energies for $v_t = 1$, $K_a = 1$ levels of He$_2$-OCS (in cm$^{-1}$).

| $J$ $K_a$ $K_c$ | Ground | $v_1$ | $v_1 + v_3$ |
|---|---|---|---|
| origin | 0.0 | 2062.4210 | 2918.4552 |
| 1 1 1 | 0.3152 | 0.3140 | 0.3132 |
| 2 1 2 | 0.7915 | 0.7884 | 0.7865 |
| 3 1 3 | 1.5046 | 1.4988 | 1.4951 |
| 4 1 4 | 2.4519 | 2.4424 | 2.4364 |
| 5 1 5 | 3.6243 | 3.6103 | 3.6013 |
| 1 1 0 | 0.3236 | 0.3224 | 0.3216 |
| 2 1 1 | 0.8137 | 0.8106 | 0.8086 |
| 3 1 2 | 1.5442 | 1.5382 | 1.5344 |
| 4 1 3 | 2.5097 | 2.4999 | 2.4937 |
| 5 1 4 | 3.7034 | 3.6890 | 3.6798 |

Note: The energy of the lowest (1 1 1) level is arbitrarily set at the calculated value of 0.3152 cm$^{-1}$, which is relative to the forbidden (0 0 0) level of the $v_t = 1$ state. The excited state energies are relative to their respective origins.

Table A-12. Quasi-experimental energies for $v_t = 1$, $K_a = 1$ levels of He$_2$-O$^{13}$CS (in cm$^{-1}$).

| $J\ K_a\ K_c$ | Ground | $v_1$ |
|---|---|---|
| origin | 0.0 | 2009.4421 |
| 1 1 1 | 0.3149 | 0.3135 |
| 2 1 2 | 0.7906 | 0.7870 |
| 3 1 3 | 1.5049 | 1.4980 |
| 4 1 4 | 2.4523 | 2.4412 |
| 5 1 5 | 3.6221 | 3.6057 |
| 1 1 0 | 0.3230 | 0.3215 |
| 2 1 1 | 0.8125 | 0.8088 |
| 3 1 2 | 1.5433 | 1.5363 |
| 4 1 3 | 2.5086 | 2.4972 |
| 5 1 4 | 3.7018 | 3.6850 |

Note: The energy of the lowest (1 1 1) level is arbitrarily set at 0.3149 cm$^{-1}$ by scaling the calculated value of 0.3152 cm$^{-1}$ for the normal isotopologue. It is relative to the forbidden (0 0 0) level of the $v_t = 1$ state. The excited state energies are relative to the quoted origin. The scaling factor relating the ground and excited states was found to be 0.9955, close to the value of 0.9961 found for the normal isotopologue.

Table A-13. Quasi-experimental energies for $v_t = 1$ $K_a = 1$ levels of $He_2\text{-}OC^{34}S$ (in cm$^{-1}$).

| $J$ $K_a$ $K_c$ | Ground | $v_1$ |
|---|---|---|
| origin | 0.0 | 2061.6649 |
| 1 1 1 | 0.3098 | 0.3088 |
| 2 1 2 | 0.7767 | 0.7741 |
| 3 1 3 | 1.4758 | 1.4709 |
| 4 1 4 | 2.4049 | 2.3968 |
| 5 1 5 | 3.5558 | 3.5439 |
| 1 1 0 | 0.3174 | 0.3163 |
| 2 1 1 | 0.7975 | 0.7948 |
| 3 1 2 | 1.5131 | 1.5080 |
| 4 1 3 | 2.4608 | 2.4525 |
| 5 1 4 | 3.6357 | 3.6235 |

Note: The energy of the lowest (1 1 1) level is arbitrarily set at 0.3098 cm$^{-1}$ by scaling the calculated value of 0.3152 cm$^{-1}$ for the normal isotopologue. It is relative to the forbidden (0 0 0) level of the $v_t = 1$ state. The excited state energies are relative to the quoted origin. The scaling factor relating the ground and excited states was found to be 0.9966, close to the value of 0.9961 found for the normal isotopologue.